\documentclass[review]{elsarticle}
\journal{Parallel Computing journal}

\usepackage{amsmath}
\usepackage{amssymb}
\usepackage{graphicx}
\usepackage{graphics}
\usepackage{color}
\definecolor{darkgreen}{rgb}{0,.5,0}
\definecolor{orange}{rgb}{1,0.5,0}
\usepackage{listings}
\usepackage{caption}
\usepackage{subcaption}
\usepackage{algorithm2e}
\usepackage{braket}
\usepackage{listings}
\usepackage{hyperref}

\begin{document}

\begin{frontmatter}

\title{ScaffCC: Scalable Compilation and Analysis of Quantum Programs
}

\author[princetonaddress]{Ali JavadiAbhari}
\ead{ajavadia@princeton.edu}

\author[princetonaddress]{Shruti Patil}

\author[ucsbaddress]{Daniel Kudrow}

\author[ucsbaddress]{Jeff Heckey}

\author[ibmaddress]{Alexey Lvov}

\author[ucsbaddress]{Frederic T. Chong}

\author[princetonaddress]{Margaret Martonosi}

\address[princetonaddress]{Princeton University}
\address[ucsbaddress]{University of California, Santa Barbara}
\address[ibmaddress]{IBM T. J. Watson}

\begin{abstract}
We present ScaffCC, a scalable compilation and analysis framework based on LLVM \cite{LLVM}, 
which can be used for compiling quantum computing applications at the logical level. 
Drawing upon mature compiler technologies, we discuss similarities and differences between compilation
 of classical and quantum programs, and adapt our methods to optimizing the compilation time and output
 for the quantum case. Our work also integrates a reversible-logic synthesis tool in the compiler to facilitate coding of quantum circuits.
 Lastly, we present some useful quantum program analysis scenarios and discuss their implications, specifically with an 
 elaborate discussion of timing analysis for critical path estimation.
 Our work focuses on bridging the gap between
 high-level quantum algorithm specifications and low-level physical implementations, while providing good scalability 
 to larger and more interesting problems.
\end{abstract}


\end{frontmatter}

\section{Introduction}\label{sec:Intro}
Quantum computing offers the possibility of efficiently solving problems which are computationally very difficult to solve using classical algorithms on classical computers. Examples of such problems include factoring of large numbers into prime factors and simulating chemical atomic systems \cite{Shor,ref:gse}. There has been a significant increase in quantum computing research in recent years, however the gap between quantum algorithms of practical interest and what can be feasibly implemented still remains.

In this paper, we describe the requirements of quantum program compilation and present ScaffCC, a compiler framework with an extensible quantum program analysis toolbox. ScaffCC's approaches are designed to scale effectively to compile programs that contain trillions of operations (instructions). Most previous works have focused on designing, mapping, and scheduling hand-optimized quantum circuits for implementing small-scale quantum algorithms. 
Although \emph{quantum error correction} is likely to dominate computation in any feasible implementation of quantum algorithms, and there are indeed many efforts to optimize this stage of code translation (e.g. \cite{bravyi1998quantum, fowler2012surface}), we have designed ScaffCC for the logical level of quantum computation (i.e. before error correction). This is because any optimization at the logical level will have a mulitiplicative effect on the required amount of error correction and the ultimate resource consumption.

Overall, this paper makes the following contributions: 
\begin{enumerate}

\item
We identify some key differences between classical and 
quantum compilation.  For example, quantum programs are a static description of quantum circuits, and are therefore specialized to certain problem parameters. As a result, they 
yield statically analyzable code, mitigating the need for optimizations 
such as branch prediction and emphasizing other optimizations such 
as parallelization of operations.
Moreover, this creates opportunities
for aggressive constant propagation and deep optimization, while
simultaneously putting greater pressure on the scalability of the compiler
algorithms employed.

\item
We present compiler algorithms and compiler output formats that
can accommodate the large scale and deep optimization found in 
our quantum benchmarks.  In particular, we find that output modularity
and a dynamic, instrumentation-driven compilation technique are important to 
managing scale. This is in contrast to conventional compiler code generation approaches which use multiple compile-time passes for optimizing and emitting code.

\item 
Despite the differences inherent in quantum compilation as opposed to
the classical case, we show the applicability of known classical
compiler algorithms, such as loop unrolling and procedure cloning, to
the domain of quantum computing.  Our compiler leverages mature
compiler technologies through the LLVM framework.

\item
We present data-flow analysis as an example of classical techniques employed in the quantum domain. In particular, we propose the use of data-flow analysis techniques, both for important \emph{program checks} such as ``no-cloning" and ``entanglements,"  and also for obtaining \emph{circuit estimates} such as the critical circuit path or its usage of qubits and operations. These metrics help focus further optimizations.

\item
We demonstrate the trade-off between accuracy and speed in analyzing the critical path length of large quantum programs. We propose three methods of increasing complexity for critical path length analysis, and discuss the scenarios in which they can help achieve better speed and accuracy.

\item 
Finally, recognizing the difficulty of hand coding math library functions in quantum programs, we observe the need for using classical reversible logic to describe sub-circuits of a quantum circuit, and hence present a novel technique for the compilation and simulation of such modules.

\end{enumerate}

The rest of this paper is organized as follows:  Sections 2 and 3 give background on quantum computation, and then an overview of the compiler we have developed to translate from high-level quantum algorithms to lower-level quantum assembly operations.  Sections 4, 5, and 6 describe the research challenges in different parts of the compiler toolflow, including techniques to manage large scale and to synthesize from classical reversible logic.  Section 7 discusses analysis passes enabled by the ScaffCC functionality. Finally, Section 8 presents related work, and Section 9 offers  conclusions.

\section{Quantum Computation}\label{sec:Background}
This section offers a brief background on basic concepts in quantum computation.

{\bf{Quantum States and Superposition:}} While classical bits exist in only one of the binary states at any given time, quantum bits, or \emph{qubits}, can exist in a \emph{superposition} state, which is a linear combination of the $\ket{0}$ and $\ket{1}$ states. This extends to multiple qubits, i.e. a quantum mechanical system with 2 qubits can be simultaneously representative of the four states $\ket{00}$, $\ket{01}$, $\ket{10}$ and $\ket{11}$. Quantum \emph{operations} can modify such superposition states simultaneously, allowing some quantum algorithms to perform faster than their classical counterparts. Quantum states also exhibit other properties such as \emph{entanglement}, which causes the state of two qubits to be dependent on each other, and \emph{no-cloning}, which restricts copying of one arbitrary quantum state into another.
 
Though a quantum algorithm uses quantum bits and operations during the computation, it must, in the end, provide a classical answer to a classical inquiry. This is achieved using \emph{measurement}, which causes a qubit to lose its superposition and collapse into a deterministic state of $\ket{0}$ or $\ket{1}$. Since this process is probabilistic in nature, quantum algorithms seek to manipulate quantum states so as to increase the likelihood of measuring the desired answer in the end.

{\bf{Quantum Operations and Reversibility:}} Any valid operation on quantum states must be \emph{unitary}. 
This implies that all operations, and in fact the entire quantum circuit, must be reversible. Analogous to classical logic gates, the quantum operations which form basic building blocks of quantum circuits are known as \emph{quantum gates}. Quantum algorithms typically describe a quantum \emph{circuit} defining the evolution of multiple qubits using basic quantum gates.  

{\bf{Compiler Implications:}} This theoretical background guides the design of an effective quantum compiler. Some of the described quantum phenomena such as entanglement between states of qubits or impossibility of copying states are important in detecting possible logical flaws in a program. Section \ref{sec:Analysis} shows how entanglement relationships by the compiler in order to inform the programmer about possible coding errors.

The reversibility criterion is also important to compilers of quantum programs; non-reversible sub-circuits need to be detected, or made reversible, for valid quantum circuit generation. In this, the compiler must be aware of the cost of qubits as the most expensive resources.

\section{Overview of ScaffCC}\label{sec:CompilerOverview}
ScaffCC compiles a program written in the Scaffold programming language, and outputs a quantum assembly (QASM) representation. It targets {\it logical} quantum computation, that is, compilation, analysis and optimizations before synthesis into machine-dependent physical-level operations.
This section gives a broad overview of the input and output languages, and the design of the ScaffCC compiler.

\subsection{Scaffold Quantum Programming Language} \label{subsec:Scaffold}
Scaffold \cite{Scaffold} is a high-level, imperative quantum programming language, designed as an extension to C. 
Scaffold includes new data types, \emph{qbit} and \emph{cbit}, corresponding to quantum bits, and classical bits obtained as a result of measurement, respectively. Furthermore, it includes basic quantum operations (gates) such as Pauli X, Hadamard, Toffoli, Rotation, etc. as built-in entities. A Scaffold program can be regarded as being composed of two parts: the quantum part containing descriptions of quantum bits and operations, and the classical part containing classical control around those operations, such as loops and conditionals.  

Similar to a C program or a Verilog classical circuit, almost every Scaffold quantum code has a hierarchical structure and is organized into \emph{modules}. Each module represents a sub-circuit of the overall program circuit, and can be instantiated within larger (parent) modules. Since quantum circuits must be ``reversible", each module must either be specified using unitary quantum operations, or be transformed as such by the compiler. Scaffold includes a class of modules novel among quantum compilers, called Classical-To-Quantum-Gate (CTQG). These allow sub-circuits to be defined as classical logical circuits. ScaffCC converts these into valid quantum codes, as discussed in Section \ref{sec:CTQG}.

\subsection{QASM Assembly Language}\label{subsec:QASM}
The quantum assembly language of QASM, proposed in  \cite{MikenIke, Svore}, describes quantum programs using a set of low level quantum gates. QASM specifies logical qubits and the sequence of gate operations performed on them. Basic data types in QASM are \emph{qbit} and \emph{cbit}, and the instruction set includes a universal set of gates (Controlled-NOT (CNOT), Hadamard (H), Phase (S), $\pi/8$ Rotation (T)), plus operations for measurement and preparation in the states $\ket{0}$ and $\ket{1}$. QASM is independent of the underlying quantum technologies, and assumes that the hardware can implement the described circuit using suitable gate transformations and error correction in the next stages of synthesis.

QASM has been used to implement and study quantum circuits for small problems using a flat circuit format~\cite{Elhoushi, MetodiCompiler, qasm}.
However, realistic quantum circuits that we examined contain between $10^{7}$ and $10^{12}$ gates, rendering full flattening infeasible. In Section 4, we introduce modifications to the original flat format that retain scalability by enabling more manageable target QASM sizes. 

\subsection{Internal Structure of the Compiler}\label{subsec:CompilerSteps}
\begin{figure*}
   \centering
   \includegraphics[width=\textwidth]{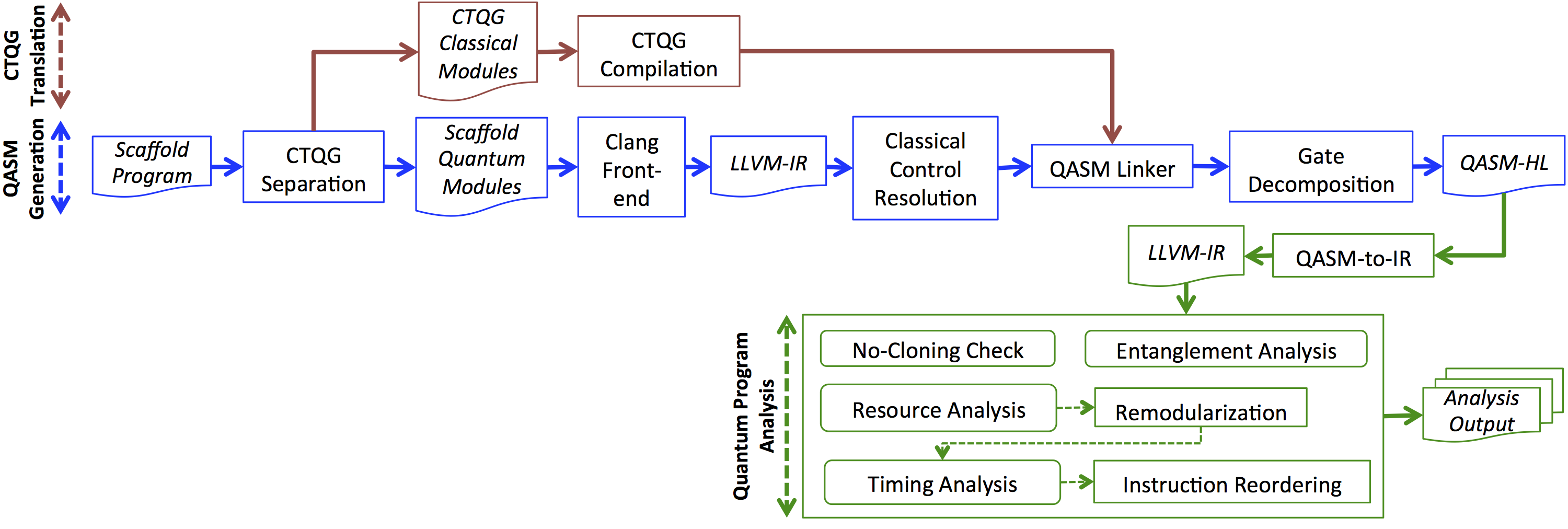}
   \caption{Internal structure of the ScaffCC compiler: The top, middle and bottom parts respectively show translation of CTQG modules (Section \ref{sec:CTQG}), QASM code generation (Section \ref{sec:GenQASM}), and quantum program analysis (Section \ref{sec:Analysis}).}
   \label{fig:compiler_steps}
\end{figure*}

Fig.~\ref{fig:compiler_steps} depicts a block diagram of ScaffCC's internal structure. We have implemented ScaffCC in \emph{LLVM}~\cite{LLVM}, a rich, open-source library of compiler technologies, by adding \emph{intrinsic functions} representative of quantum gates and a datatype representative of qubits. Furthermore, we have extended \emph{Clang}, a C-family front-end to LLVM, to accommodate parsing of our language. 

The first step of compilation is to separate the modules in the program which are marked as CTQG. These modules have been defined by the programmer using classical gates, and are handled by the separate CTQG sub-compiler as described in Section \ref{sec:CTQG}. CTQG's output is translated directly to QASM without going to LLVM's intermediate format, and is linked with the output of the quantum modules after they have been converted to QASM. Although this approach yields fast output code generation, it is not suitable for whole program analysis since a part of the code will bypass the LLVM-IR representation. Thus, we have implemented a \emph{QASM-to-IR} translator which we use to convert the entire program once it has been compiled. This provides correct input for quantum program analysis.

A critical code generation issue lies in the degree to which output code can or should be linearized (or flattened). We refer to this as ``classical control resolution". Our goal is to establish a judicious balance---we wish to flatten as much as possible in order to support efficient synthesis of quantum circuits, while also keeping enough abstraction to ensure circuit generation remains tractable. During the compilation of non-CTQG modules, it becomes necessary to process some of the classical instructions within them, in order to remove high-level abstractions and obtain sub-circuits that clearly specify the sequence of gate operations and qubits which are acted upon. This amounts to flattening the program on a per-module basis, and is required for correct scheduling and mapping during later stages of the toolchain. Unfortunately, performing code linearization in a way that scales well and does not result in a time and space explosion is non-trivial. Section \ref{sec:GenQASM} has a detailed description of this step and explores ways to make it execute faster.

The final phase of the compiler performs a decomposition of unitary operations into supported gates in QASM, which is a subset of those allowed in Scaffold. This is a key step in the translation of a high-level program into a standard assembly language, and is similar to instruction selection in classical compilers. For some gates, this is a straight-forward process. For example, the output of CTQG contains many ``Toffoli" operations, which in order to be compatible with QASM, would each be substituted by a fixed 16-gate sub-circuit. Other gates, such as rotations by arbitrary angles, may be more complex. We employ a state-of-the-art method, as proposed in \cite{Kudrow}, to approximate these gates.

Finally, as Section \ref{sec:Analysis} discusses in detail, ScaffCC can perform a range of useful analyses on its input programs, both for program correctness checks and for circuit estimates. The LLVM toolkit represents computations as graphs, which facilitates program analysis.

\subsection{Scaffold Benchmarks}\label{sec:benchmarks}
We perform a comprehensive study of the performance of our compilation and analysis techniques using a set of eight quantum algorithms. The coding of these benchmarks and our tools originally began in the IARPA quantum computer science program. These benchmarks cover many common themes in quantum algorithm design: Quantum Fourier Transform, Classical Oracles, State Distillation, Random Walk, and Amplitude Amplification among others. This constitutes one of the first studies in compiling quantum programs of this large size and broad scope.

1) \emph{Grover's Search Algorithm}: Uses quantum amplitude amplification to search a database of $2^{n}$ entries. It is parameterized by \emph{n} (log of the number of entries)~\cite{Grover}.

2) \emph{Binary Welded Tree (BWT)}: Uses quantum random walk to find a path between an entry and exit node of a binary welded tree. The benchmark is parameterized by height of the tree (\emph{n}) and a time parameter (\emph{s})~\cite{ref:bwt}.

3) \emph{Ground State Estimation (GSE)}: Uses quantum phase estimation to estimate the ground state energy of a molecule. The benchmark is parameterized here by the molecular weight (\emph{M}), but could also be parameterized by precision~\cite{ref:gse}.

4) \emph{Triangle Finding Problem (TFP)}: A quantum algorithm to find a triangle within a dense, undirected graph using quantum random walk. The program is parameterized by the number of nodes \emph{n} in the graph~\cite{ref:tfp}.

5) \emph{Boolean Formula (BF)}: Computes a winning strategy for the game of Hex with quantum random walk. The benchmark is parameterized by size of the Hex board \emph{(x, y)}~\cite{ref:boolean_formula}.

6) \emph{Class Number (CN)}: A problem from computational algebraic number theory that uses Quantum Fourier Transform to compute the class group of a real quadratic number field. The program is parameterized by \emph{p}, the number of digits after the radix point for floating point numbers used in computation~\cite{ref:class_number}.

7) \emph{Shor's Factoring Algorithm}: Performs factorization using the Quantum
  Fourier Transform \cite{Shor}. The benchmark is parameterized by $n$, the
  size in bits of the number to factor.

8) \emph{Secure Hash Algorithm-1 (SHA-1)}: A quantum implementation of the classical
  algorithm \cite{ref:sha1}. The benchmark is parameterized by the size of the
  message in bits ($n$).

\section{Managing Scalability Through Choice of QASM Format}\label{sec:QASM}
As stated before, an important research issue concerns managing the scale of generated QASM code in large-scale benchmarks. Therefore, here we consider QASM format adjustments over previous flat-code proposals, and study their impact on code generation feasibility.

\begin{figure}
\centering
\begin{subfigure}[b]{0.45\linewidth}
\centering
\begin{lstlisting}
#define n 1000
module foo(qbit q[n])
{
  for(int i=0;i<n;i++)
    H(q[i]);   
  CNOT(q[n-1],q[0]);
}
module main()
{
  qbit b[n];
  foo(b); 
}
\end{lstlisting}
\caption{Scaffold}
\label{fig:scaffold_format}
\end{subfigure}
\hfill
\begin{subfigure}[b]{0.45\linewidth}
\centering
\begin{lstlisting}
qbit b[1000];
H ( b[0] );
H ( b[1] );
.
.
H ( b[999] );
CNOT ( b[999] , b[0] );
\end{lstlisting}
\caption{QASM-F format}
\label{fig:qasmf_format}
\end{subfigure}
\hfill
\begin{subfigure}[b]{0.45\linewidth}
\centering
\begin{lstlisting}
module foo ( qbit* q )
{
  H ( q[0] );
  H ( q[1] );
  .
  .
  H ( q[999] );
  CNOT ( q[999] , q[0] );
}
module main (  )
{
  qbit b[1000];
  foo ( b );
}
\end{lstlisting}
\caption{QASM-H format}
\label{fig:qasmh_format}
\end{subfigure}
\hfill
\begin{subfigure}[b]{0.45\linewidth}
\centering
\begin{lstlisting}
module foo ( qbit* q )
{
  H ( q[0:999] );
  CNOT ( q[999] , q[0] );
}
module main (  )
{
  qbit b[1000];
  foo ( b );
}
\end{lstlisting}
\caption{QASM-HL format}
\label{fig:qasmhl_format}
\end{subfigure}
\caption{Code Snippets for QASM-F, QASM-H and QASM-HL: Progressively more classical control is retained. Note that Scaffold does not contain pointers or allow their manipulation, but QASM address representation for accessing memory resembles C syntax for ease of use with LLVM.}
\label{fig:example_qasm_formats}
\end{figure}

{\bf{Hierarchical QASM format (QASM-H):}} Similar to hardware description language formats, QASM programs can be represented by a space-consuming \emph{flat} description, or by a denser \emph{hierarchical} description which takes advantage of sub-circuit duplications to reduce the output code size. Some modularity is also desirable for program analysis of large codes. Analysis techniques when applied hierarchically reduce analysis time and memory usage, thus scaling better to large program sizes. We demonstrate this through the example of \emph{timing analysis} in Section \ref{subsec:Timing}.

{\bf{Hierarchical QASM with Loops (QASM-HL):}} Further information about repeating quantum operations can be retained within the QASM format, in the form of loops. Quantum circuits show two prominent types of quantum operations: The first type are operations that are applied to a large set of qubits. These are used, for example, when transforming qubits prepared in the ground state into initial superposition states. Due to the absence of qubit dependencies, these operations are highly parallel and are implemented simultaneously when the hardware technology allows it. (For example one can use control technologies such as microwave traps that affect a large number of qubits at the same time.) We denote these as \emph{forall} loops.

The second type of operations are serially repeated transformations, typically used in quantum algorithms to converge to a more precise solution. For example, Grover's Search Algorithm makes use of a repeated invert-and-reflect operation that gradually increases the likelihood of measuring the correct answer.  In the physical implementation, the control exercised for the sequence of operations within the loop body can be synthesized once, and then reused. We denote these as \emph{repeat} loops.

In order to identify quantum \emph{forall} and \emph{repeat} loops in high-level programs, we define a \emph{pure} quantum block as a basic block that conforms to the following criteria: 1. A pure quantum block does not contain classical computation instructions such as arithmetic or compare instructions; 2. It does not contain function calls which have non-quantum data types as arguments; 3. In a pure quantum block the qubit array variables depend directly on the loop induction variable. Through static analysis of the loops around the purely quantum blocks, we can obtain trip counts to provide the number of repetitions for the repeat loops, and loop values to provide the range of qubits that are simultaneously operated upon in the forall loops.  This allows for efficient optimizations and analyses.

\begin{figure}
   \centering
   \includegraphics[width=\linewidth]{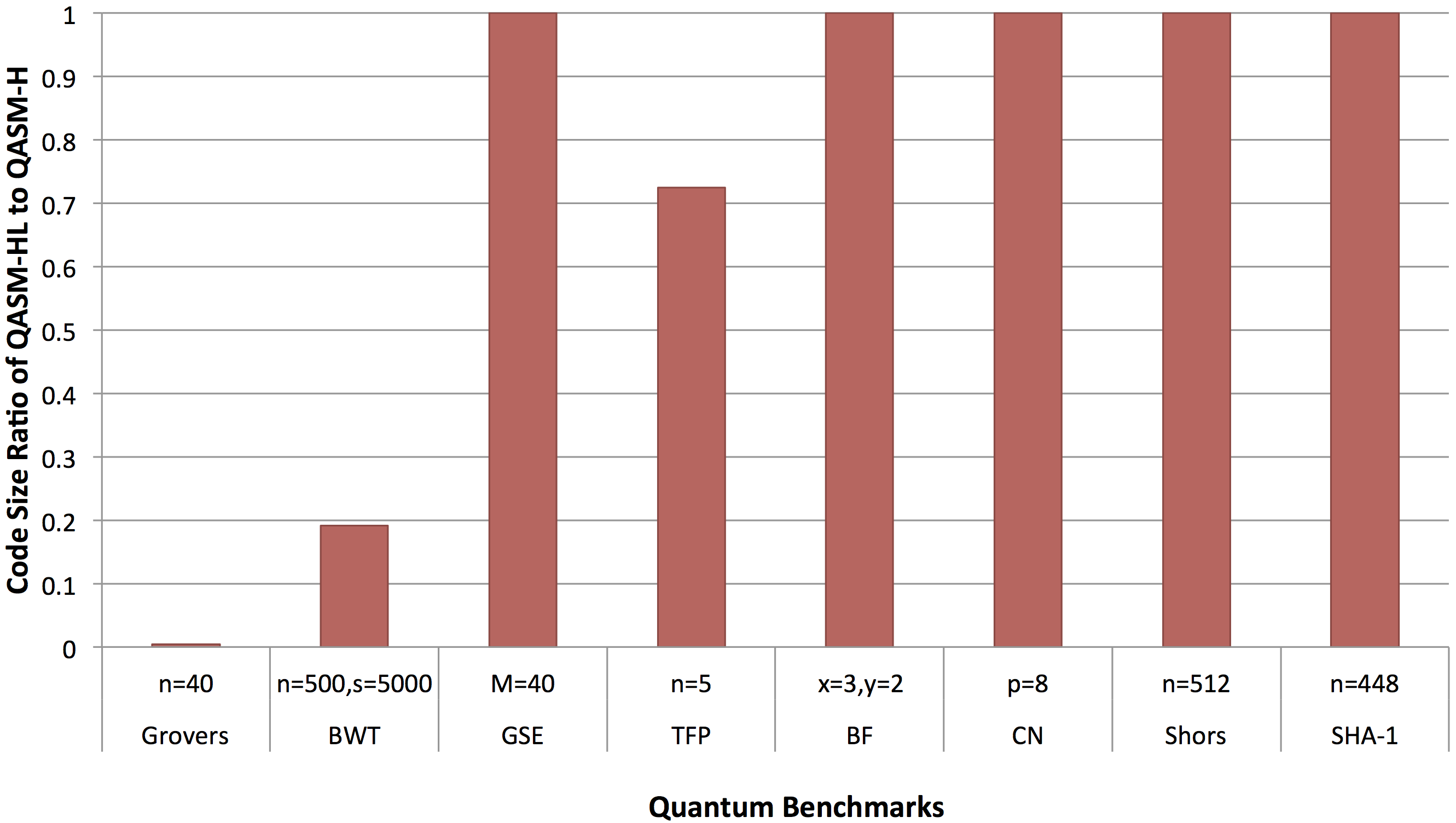} 
   \caption{Reduction in code size of QASM-HL compared to QASM-H output, due to retention of quantum loops.}
   \label{fig:qasm_hl_codesize}
\end{figure}

{\bf{QASM Code Size Comparison:}} Fig.~\ref{fig:qasm_hl_codesize} shows the reduction in code size when using QASM-HL over QASM-H. A great advantage in code size ($\sim$200,000X on average) is already obtained across all benchmarks when using QASM-H as opposed to flat QASM.

Referring to this figure, QASM-HL output format particularly improves code size for the Grovers and BWT algorithms, making an exponential growth with problem parameters into a linear one. The reason is that these algorithms make use of \emph{repeat} blocks with high iteration count, in a manner that converges the quantum states to the correct results. As programs scale, the increased number of quantum operations is captured within the \emph{repeat} loop of QASM, keeping the resulting QASM sizes small. On the other hand, the TFP algorithm has numerous \emph{forall} blocks, but a relatively low number of \emph{repeat} blocks. As the problem size for this algorithm scales, the trip counts of \emph{forall} loops capture the increased number of qubits being operated upon, resulting in some code improvement. For three of the benchmarks, not much advantage is gained when using QASM-HL over QASM-H. In the GSE and Shor's programs, very few pure quantum loops and with low trip counts exist, impeding the effectiveness of loop retention. In addition, a major part of the BF, CN and SHA-1 circuits are compiled using the CTQG sub-compiler, which outputs a flat circuit format. Quantum loops constitute a very small percentage of the non-CTQG part, resulting in only slight code size improvements. Overall, QASM-HL's advantage is in making compilation tractable for more programs.

\section{Code Generation and Scaling}\label{sec:GenQASM}
Another important goal of ScaffCC is to scale well with increasing circuit sizes. As previously defined, QASM-HL supports this by allowing modularity and repetitions in the output code, which mitigates the size explosion that results from flattening the whole circuit. However, with the exception of some loops, QASM-HL still requires per-module flat code to enable effective circuit synthesis. Therefore, many classical control constructs, such as if-then-else conditionals, non-quantum loops, parameterized modules, etc.~must be processed in the compiler. Scaffold programs contain the description of a quantum circuit and are thus specialized for a particular set of input parameters (or problem sizes), yielding deeply analyzable programs. This fixed-trace nature of program control-flow and its non-dependence on qubit states means that all classical control-flow constructs can be resolved in the compiler. 

This section begins with a motivating example regarding the need for classical control resolution, and then describes methods for compiler implementations of it. The speed and tractability advantages of our second method over the first are discussed at the end.

Consider Fig.~\ref{fig:control_resolution} which shows a segment of a Scaffold program where the module \verb$main$ contains calls to module \verb$Oracle$ located inside two nested loops. For each different value of \verb$j$, a different \emph{version} of \verb$Oracle$ is called, since the rotation angle in the \verb$Rz$ rotation gate changes. In order to correctly decompose this gate, the compiler needs to disambiguate these different module versions, and obtain the correct rotation angle for each one to arrive at its equivalent set of gates. This is why, for example, QASM-HL does not contain parameterized modules. We investigate how the compiler can automate this resolution of classical control.

\begin{figure}[h]
   \centering
	\begin{lstlisting}
#define s_ 3000      // iteration count	

module Oracle (qbit a[1], qbit b[1], int j) {
  double theta = (-1)*pow(2.0, j)/100;
  X(a[0]);
  Rz(b[0], theta);
}

module main () {
  qbit a[1], b[1];
  int i, j;
  for (i=1; i<=s_; i++) {
    for (j=0; j<=3; j++) {
      Oracle(a, b, j);
    }
  }	
}
	\end{lstlisting}
   \caption{Example Scaffold program showing the need for classical control resolution. Different versions of the same module with different gate sets are created, but can be discovered either statically using compiler passes such as loop unrolling and procedure cloning, or dynamically using instrumentation and execution.}
   \label{fig:control_resolution}
\end{figure}

\subsection{Pass-Driven Approach}\label{subsec:PassDriven}
Our first approach, \emph{pass-driven}, relies on static usage of transformation and analysis passes such as heavy constant propagation and constant folding.  It processes the modules in the call graph of the program in depth-first, pre-order and unrolls all loops that have not been marked as quantum loops. It further clones those modules that are called with different parameters in multiple call-sites, and uses inter-procedural constant propagation to specialize those modules. These steps are repeated until there is no further action to be taken. Since ScaffCC uses the LLVM infrastructure, several pre-written passes are available for these transformations; we adopt these and expand on them. 

Referring back to  Fig.~\ref{fig:control_resolution}, we begin by unrolling the inner loop in module \emph{main} by a factor of 4, which causes all call sites to module \verb$Oracle$ to have constant call parameters. We then use procedure cloning to create a module clone from each call site that has a unique set of input parameters. We then use inter-procedural constant propagation to propagate the input parameter constants of the call sites into each corresponding module. Repetitive application of these transformation passes (loop unrolling, function cloning and constant propagation) yields a program that preserves modularity but is flattened on a per-module basis. This process is equivalent to a \emph{partial execution} of the code. This example also illustrates a further optimization: simple loops, as discussed in Section \ref{sec:QASM}, may be kept since their loop bodies are all quantum and their resource usage can be multiplied by the loop trip count. The outer loop in module \emph{main} is an example of this kind of loop, where the input parameter \emph{``s"} is a timestep variable that indicates the number of required iterations over a circuit segment in order to converge to the answer. This is the major source of computation in this benchmark; avoiding unrolling the loop offers a great gain in space complexity.

\subsection{Instrumentation-Driven Approach} \label{subsec:InstDriven}
The pass-driven approach can quickly become cumbersome for algorithms that have many large modules. The transformation passes create large intermediate code sizes, with unacceptable space and time costs. To address this, the \emph{instrumentation-driven} approach shifts from static code transformation to an execution-based transformation. This shifts the job of resolving classical dependencies to the classical processor, recognizing that fast classical processors can be used to execute through classical portions of the code and collect information regarding the quantum part.

For such source-to-source rewriting, a naive instrumentation approach would be to instrument the quantum instructions to print themselves textually as the classical component executes. However this will result in a flat output. For QASM-HL output, the instrumentation approach must preserve modularity during execution. For this purpose, the program is modified to execute in two modes, the quantum mode and the classical mode. In the quantum mode, the instructions in a module are rewritten into the quantum assembly format, while in the classical mode, the call paths are followed to determine the next set of modules to be translated. In particular, we use ``procedure cloning" to create the quantum version of the module from the original version (denoted as the classical version).

In the quantum mode, each module resolves the sequence of its quantum operations and quantum data references, by executing the classical control instructions within it. Once the quantum operations and their operands are extracted, they are converted into QASM-HL format and written to the output file. To achieve this, each quantum operation is instrumented with a print function that prints the operation type and the resolved data operand references in the QASM-HL syntax. The function calls to other modules are also instrumented, but removed in order to prevent them from executing. Once the instrumentation pass is performed, a dead code elimination pass is used to remove the dead instructions in the quantum version.

The classical version of a module is instrumented to invoke the quantum version of the module, before executing the function calls contained within it. In this module, the quantum instructions are removed, leaving only the function calls intact. To prevent repeated execution of the module, runtime decision instructions are added at the beginning of the classical version. We use the technique of \emph{memoization} to determine if the module was executed previously, by inserting it into a look-up table. As further optimization, loop iterations that have exactly identical call sequences, including their call parameters, are removed so that the call sequence is executed only once.

\subsection{Compilation Speed Comparison}
\begin{figure}
   \centering
   \includegraphics[width=\linewidth]{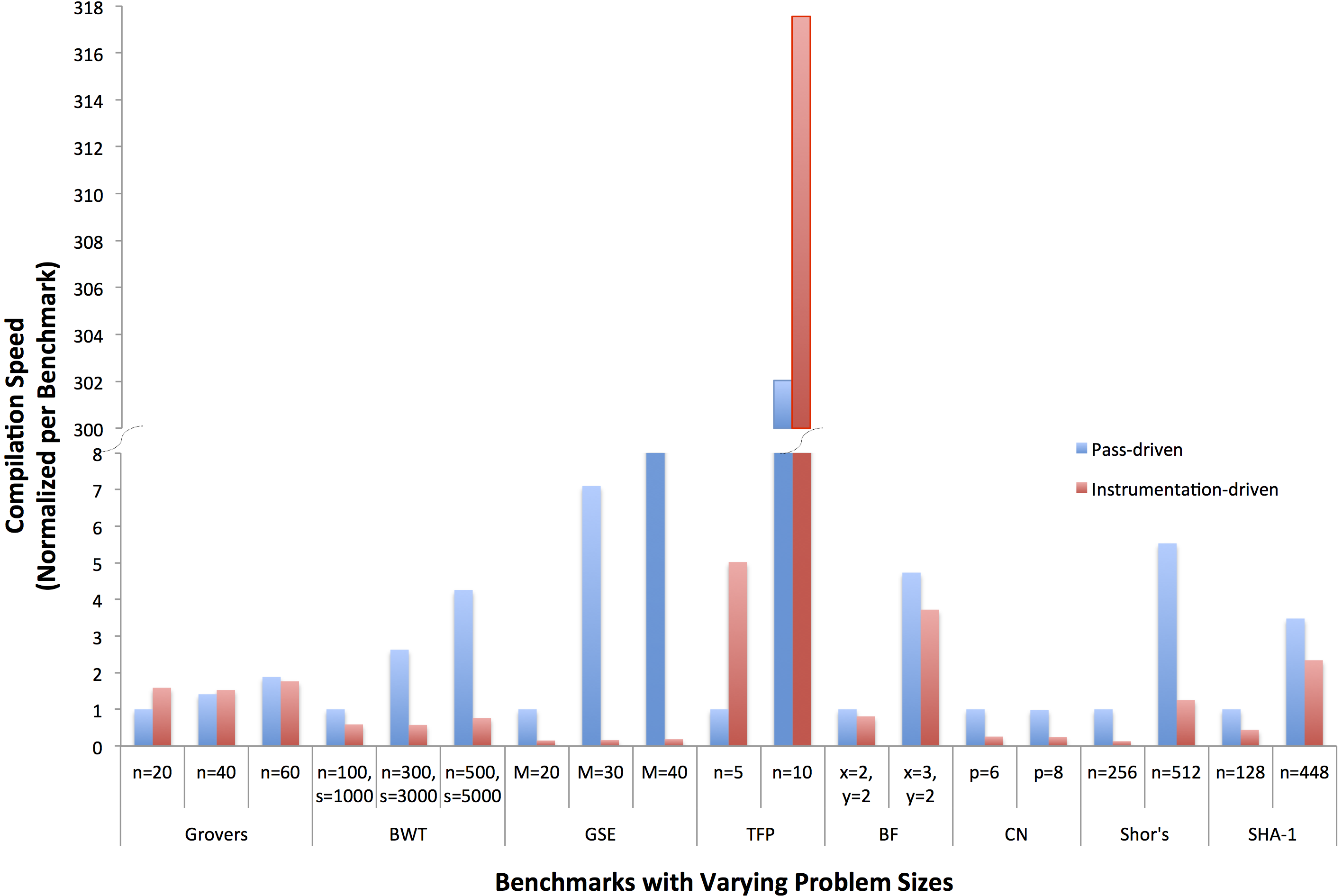}
   \caption{Improvement in compilation time with the instrumentation-driven technique over the pass-driven, for different problem sizes. Figure is scaled to the pass-driven time for the smallest problem size. Pass-driven compilation can be faster for small benchmarks, while instrumentation-driven compilation supports larger benchmarks (scales better). On average, instrumentation-driven technique is 3X faster.}
   \label{fig:compilation_time_comparison}
\end{figure}

Fig.~\ref{fig:compilation_time_comparison} shows the improvements of the instrumentation-driven approach over the pass-driven approach in overall compilation time across the range of all quantum benchmarks. Results were collected using a 2.27 GHz, Intel Xeon CPU with 24 MB of shared cache and 126 GB of RAM. For each benchmark, the compilation time is normalized to the pass-driven time of the smallest problem size. As problem sizes increase, the instrumentation-driven approach scales better than the pass-driven approach. This amounts to significant improvements in compilation time for large benchmarks. For example, for the Triangle Finding Problem with problem size $n=15$, the instrumentation-driven approach generates QASM-HL code within $\sim20$ hours, while compilation using the pass-driven approach takes several days.

\section{CTQG: Classical-To-Quantum-Gate Conversion} \label{sec:CTQG}
In many important quantum algorithms a large portion of
modules use only \emph{classical} reversible logic operations---operations which can be decomposed into the universal set of NOT, CNOT and Toffoli gates. These are often called ``classical oracles."
Also, unlike general quantum circuits, classical
oracles can be {\it simulated} on a conventional computer allowing a
continuous development cycle: 1. write code, 2. test by simulation, 3. 
correct bugs (if any). Compiling classical oracles separately gives an advantage of 
being able to verify a significant part of the quantum circuit by simulation.

At the first step of compilation ScaffCC detects all purely classical
reversible logic modules and compiles them using CTQG, a sub-compiler converting to flat QASM format. Later during
compilation, these precompiled classical oracles are inserted verbatim into
the final code every time a call to an oracle is encountered.  Also CTQG
allows code developers to simulate any oracle on any set of input signals for
verification and debugging purposes.

Many important basic operations such as integer arithmetic, fixed-point
arithmetic, manipulations with bit strings, allocation of ancilla signals, 
if-then-else statements and loops with non-quantum bodies can be
expressed solely by means of classical reversible logic.  CTQG uses state-of-the-art built-in algorithms to compile these operations and pass them as
QASM code to ScaffCC.

The basic integer arithmetic operations in reversible logic are $a = a +
const$, $a = a + b$, $a = a - b$ and $a = a + bc$ where the variables are
integer numbers in standard binary $n$-bit representation.  For reversible
adder and subtracter, CTQG uses a recently developed algorithm by Cuccaro et
al ~\cite{C05} which uses $6n-3$ CNOT gates, $2n-2$ Toffoli gates and does
not require any ancilla signals at all.  That is, CTQG adder and subtracter
have size linear in the bit width of the arguments.  The CTQG
integer multiplier uses similar ideas (see ~\cite{C05}); it has size
$O(n^2)$ and uses no ancilla signals either.  Using a constant
integer expression inevitably requires ancilla signals because reversible
logic does not allow constant `0' or `1' gates.  However CTQG automatically
recycles ancillas used for representation of constants that are no longer
needed.  For example only 8 ancilla signals (not 24 as with a brute force
approach) will be allocated by CTQG for the module that computes
$\{
a = a + 231\mbox{\small [11100111]};
b = b + 219\mbox{\small [11011011]};
c = c + 189\mbox{\small [10111101]};
\}$.

Fixed-point arithmetic analytic functions such as $1/x,$ $e^x,$ $\sin x,$
$\cos x$ and $\ln x$ are much harder to implement in reversible logic.  To
the best of our knowledge, there exist no purely reversible circuits
for these functions.  CTQG has a built-in implementation of these functions
which uses much fewer ancillas than a brute force Taylor series
approach.  For example for $1/x$ we use infinite product representation:
$$
1/x = 
(2 - x)\cdot
\big(1 + (1-x)^2\big)\cdot
\big(1 + (1-x)^4\big)\cdot
\big(1 + (1-x)^8\big)\cdot \ldots
$$
which has doubly exponential convergence $\forall\:x\in [1/2, 1]$,
and produces $O(n^2\ln n)$ gates and $O(n\ln n)$ ancillas.
For $e^x,$ $\sin x,$ $\cos x$ and $\ln x$ our built-in functions produce 
$O(n^3\ln n)$ gates and $O(n^2\ln n)$ ancillas.

In order to produce \ \ {\tt if (bit) \{body\}} \ \ circuits, we
add {\tt bit} as an extra control signal to every gate of {\tt \{body\}}. 
This transforms NOTs to CNOTs, CNOTs to Toffolis, Toffolis to 3-control
Toffolis, etc.  Any $n$-control Toffoli then decomposes into a number of
regular Toffolis.  Arbitrary depth embedded if-then-else decomposes into
elementary reversible gates by applying the above procedure several times.

Generally neither conventional nor reversible circuits can have loops.
However if the maximum number of loop iterations can be predetermined, then the loop can be ``unrolled'' producing an amount of gates
approximately equal to this maximum iteration count multiplied by the number of gates in the loop body. Fig.~\ref{fig:ctqg_code} is an example of a circuit written in CTQG
that computes $1 + 2 + 3 + \ldots + n$ for a given iteration count
in a brute force fashion. 

\begin{figure}[htbp]
   \centering
     \begin{lstlisting}
#define M 100

module main_ctqg(qint[16] sum, qint[16] i, qint[16] n){
  int control_i;
  $ i := 1;
  $ sum := 0;

  for (control_i = 1; control_i <= M; control_i++) {
    $if (i <= n)
      $ sum += i;
    $endif
    $ i += 1;
  }
}

	\end{lstlisting}
	\caption{Sample CTQG code, showing the usage of loops.}
	\label{fig:ctqg_code}	
\end{figure}   

Although CTQG has a large variety of highly optimized built-in reversible
logic functions and methods for generating new reversible functions from the
existing ones, they are not sufficient to generate all reversible functions
just by their conventional (non-reversible) description {\it optimally} with
respect to the usage of ancilla signals.  As shown in \cite{S03}, the
set of gates \{NOT, CNOT, TOFFOLI\} is {\it universal} for all reversible
boolean functions which represent even permutations on the set of all
possible input bit strings.  Only one reusable ancilla signal is required to
remove the ``even permutations'' constraint.  \cite{S03} gives an explicit
algorithm for representing any reversible boolean function given by a table
of values as a sequence of \{NOT, CNOT, TOFFOLI\} with only one extra
ancilla signal.  Providing that its table of values fits in memory, any
small reversible circuit can be generated this way {\it optimally} and then
be programmed in CTQG by directly listing the sequence of \{NOT, CNOT,
TOFFOLI\} gates.

CTQG is a one-pass compiler and is able to produce QASM output gate by gate
``on the fly'' without remembering any of the previously produced gates. Thus, it can work on circuits as large
as $10^{12}$ - $10^{13}$ gates, with the limiting factor being only the
runtime but not the memory size.

\section{Quantum Program Analysis} \label{sec:Analysis}
One of the most important uses of a quantum compilation framework is to obtain information about quantum algorithms and their implementation. Programming for quantum devices can be error-prone---one must have good reason to believe that the intent of the algorithm is reflected correctly, and that the implementation does not violate the laws of quantum mechanics. An example is the \emph{no-cloning theorem}, which requires that the state of one qubit cannot be copied into the state of another while maintaining the first state \cite{MikenIke}. This is a necessary, albeit not sufficient, condition on the soundness of code. As a result, ScaffCC uses aliasing analysis to emit error messages when a programmer tries to use a multi-qubit gate on the same qubit, since that quantum state cannot be mapped onto two distinct qubits.

The next sections describe ScaffCC analyses that not only help in program validity checks, but also give timing or resource estimates for the algorithm's circuit.

\subsection{Entanglement Analysis}

Entanglement is a fundamental phenomenon in quantum mechanics, denoting a logical relation between measured states of qubits. An example wave function of two entangled qubits is $|\psi^{+}\rangle = (1/\sqrt{2})|00\rangle + (1/\sqrt{2})|11\rangle$. It shows that the measurement states of the two qubits are logically related to each other. For example if one qubit is measured and collapsed to state $\ket{0}$, the other qubit will also collapse to the same state. This phenomenon is extensively used for logical transformations of quantum states and for fast communication using quantum teleportation. Further, it is the key reason behind exponential speed-up possible with certain forms of quantum computation \cite{EntanglementSpeedsUp}. 

Since entanglements affect the final outcome of qubit states, a view of entanglements occurring within a quantum program is useful to the programmer for both designing algorithms and debugging. To analyze the large number of qubits in a quantum program, we use data-flow analysis techniques to automate the process of tracking entanglements. The entanglement analysis pass in ScaffCC performs a conservative analysis, adding annotations in the output QASM-HL program to denote  possibly entangled qubits. Fig. \ref{fig:annotated_entanglements2} shows an example of a module annotated with its entanglements.

\begin{figure}[h]
   \centering
     \begin{lstlisting}
module EQxMark_1_1 ( qbit* b , qbit* t ) {
   ...
   Toffoli ( x[0] , b[1] , b[0] ); 
   // x0, b1, b0
   Toffoli ( x[1] , x[0] , b[2] ); 
   // x1, x0, b2, b1, b0
   Toffoli ( x[2] , x[1] , b[3] ); 
   // x2, x1, b3, x0, b2, b1, b0
   Toffoli ( x[3] , x[2] , b[4] ); 
   // x3, x2, b4, x1, b3, x0, b2, b1, b0
   CNOT ( t[0] , x[3] ); 
   // t0, x3, x2, b4, x1, b3, x0, b2, b1, b0
   Toffoli ( x[3] , x[2] , b[4] ); // x3
   Toffoli ( x[2] , x[1] , b[3] ); // x2
   Toffoli ( x[1] , x[0] , b[2] ); // x1
   Toffoli ( x[0] , b[1] , b[0] ); // x0
   ...
}
// Final entanglements:
// (t0, b4, b3, b2, b1, b0);
 \end{lstlisting}
   
   \caption{Entanglement annotations in the \emph{EQxMark} module of Grover's Search benchmark. Entanglements are added as a comment to every instruction that creates them.}
   \label{fig:annotated_entanglements2}
\end{figure}

Two qubits are entangled when their individual wave functions are inseparable. In reality, determination of entanglement would require precise tracking of quantum states and transformations of qubits. On the other hand, a conservative analysis without knowledge of actual states is possible by tracking simply the interactions with other qubits. It is based on the observation that if two qubits interact, they are likely to have become entangled with each other. Such interactions occur when multi-qubit operations are performed. In particular, of the primitive gates allowed by Scaffold, the CNOT and Toffoli operations potentially create entanglement among their operand qubits. ScaffCC performs this analysis for qubits in every module: \emph{control} and \emph{target} qubits from multi-qubit operations are stored in a table as they are encountered within each module, and the instructions are annotated with these pairs. Since the entanglement property is symmetric, reflexive and transitive, the previous entanglements of the control and target qubits are also added to the annotations.

In addition to compute instructions, quantum programs also contain \emph{uncompute} instructions to reverse state changes of ancilla qubits. The CNOT and Toffoli operations are inverse functions of themselves; therefore they create disentanglements when reapplied to the same set of control and target qubits. Thus, entanglement analysis also involves tracking of uncompute portions in a module. 

\begin{figure*}[h]
   \centering
   \includegraphics[width=\textwidth]{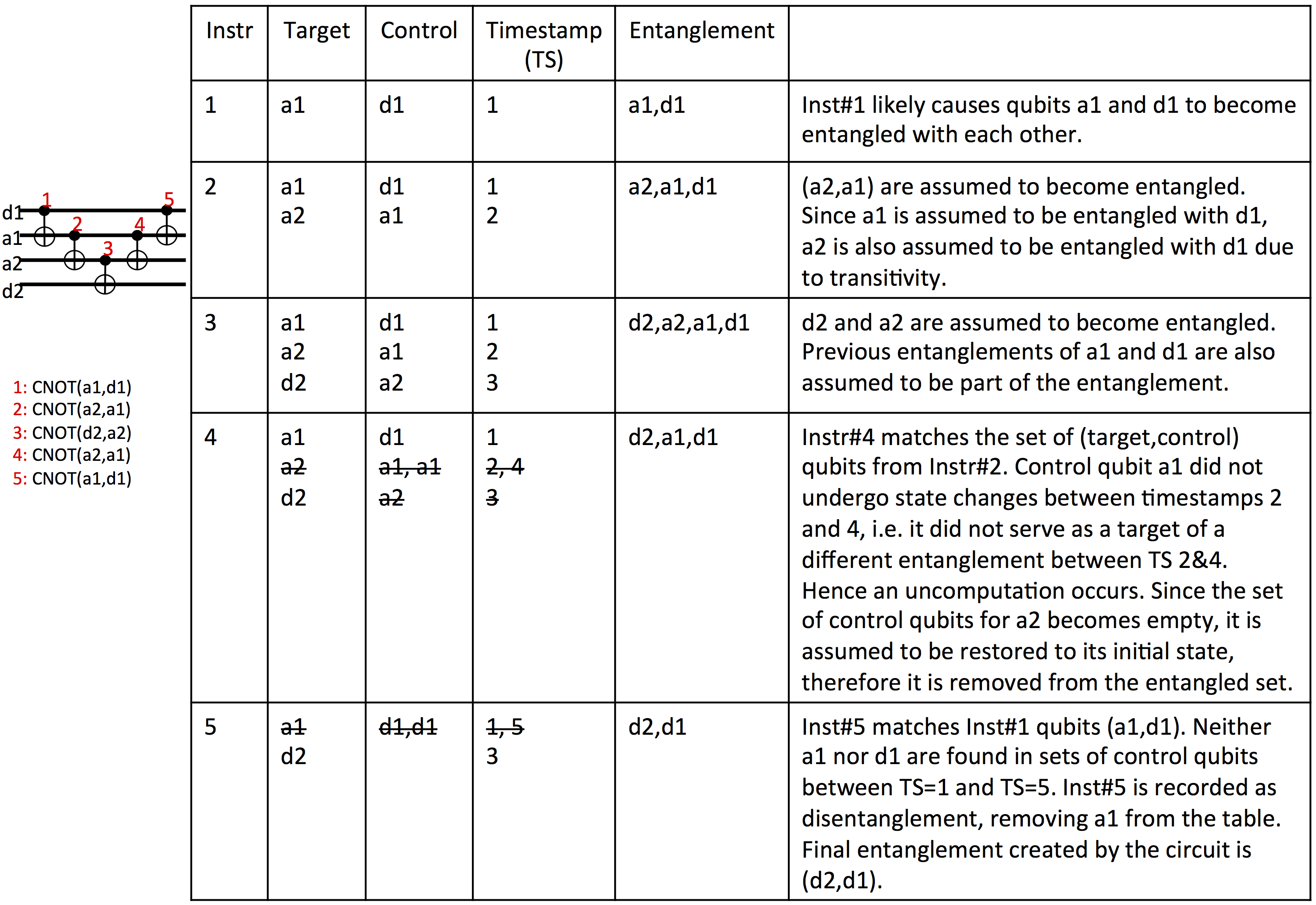}
  \caption{Example entanglement analysis of a quantum circuit that operates on two data qubits $d1$ and $d2$, using two ancilla qubits $a1$ and $a2$. The final set of entanglements realized by the circuit is $(d1,d2)$.}
   \label{fig:entanglement_analysis}
\end{figure*}

To be able to identify disentanglements, sets of control and target qubits are stored along with a timestamp for each gate. When the same gate with the same set of (target, control) qubits is re-encountered, the control qubits are examined for state changes since the timestamp of the original instruction. Changes are determined by whether the qubits served as target qubits in other instructions. If any changes were determined since the entangled instruction, the (target, control) pair is retained in the table, along with the new pair and timestamp. Otherwise, the instruction is marked as a reverse operation, and a disentanglement is recorded. This removes the (target, control) entries from the table. As a consequence, if the set of control qubits for a target qubit becomes empty, it is assumed to have been restored to its original state, and is removed from the set of entangled qubits. Fig.~\ref{fig:entanglement_analysis} illustrates this process to determine the entanglements and disentanglements for a small example circuit with two data qubits and two ancilla qubits. 

{\bf{Disentangled qubit check:}} Entanglement analysis enables the addition of an important quantum program check, which we call the \emph{disentangled qubit check}, to track abandoned qubits in a program. When ancilla qubits are not uncomputed, their wave functions stay entangled with the wave functions of data qubits. This interferes with the probabilities of measured states of data qubits, which may eventually result in incorrect outputs. To avoid these side-effects, for every module in a quantum program, each newly instantiated qubit must have been either uncomputed or measured at the end of the operation. At the end of entanglement analysis over each module, ScaffCC examines the final entanglements. On encountering a module that has remnant ancillas in the list of final entanglements, a non-uncomputed qubit warning is generated.

\subsection{Resource Analysis}\label{subsec:RE}
The high implementation cost of  qubits and operations underscores the importance of program analysis that can quickly calculate the amount of resources consumed by that program. This number can serve as an early comparison of the resource requirements of different algorithms before implementation on a physical device, as well as a form of feedback to other parts of the compiler (e.g as discussed in \ref{subsec:Timing}). Qubits remain the most expensive resources in quantum computing, but the number of gates also matters -- more gates increase the likelihood of error, thus requiring more error correction which in turn requires more qubits.

Resource analysis as a form of whole-program analysis can also be carried out using pass-driven and instrumentation-driven approaches, similar to what was discussed in Section \ref{sec:GenQASM}. In this case, either an additional compiler pass would count the number of qubits and operations on the LLVM-IR code, or instrumentation would yield a program which upon execution collects its own resources. The instrumentation-driven approach again performs better for larger problems.

Our instrumentation-driven approach is slightly different in the case of resource estimation --- quantum operations are converted into increment operators that count the occurrences of each gate on a per-module basis and add them recursively to their parent modules. However, since quantum algorithms can contain on the order of trillions of operations, it would be inefficient to traverse all operations individually. \emph{Memoization} can be used here too to exploit program modularity, but with the goal of preventing repeated same-module calls; this speeds resource analysis. This memoization requires the previously-mentioned look-up table to be expanded into a hash table that also records the counts of different resources. All hash table entries are populated on the first execution of each unique version of a quantum circuit module. For all subsequent calls, if the module and its call parameters match an entry in the table, the previously calculated results are used, without recalculation. This is possible because procedure calls in the Scaffold language do not have side-effects on the number of resources within each procedure. Table~\ref{tab:hash_table} depicts an instance of this table for the example in Fig.~\ref{fig:control_resolution}.

\begin{table}[h] 
\caption{Memoization hash-table for speeding up resource analysis for the example in Fig. \ref{fig:control_resolution}.} 
\centering 
\resizebox{0.9\linewidth}{!}{
\begin{tabular}{| c c c | ccccc |} 
\hline\hline 
 & & & \multicolumn{5}{c}{Resources} \\[-1ex]
 \raisebox{1.5ex}{Module} & \raisebox{1.5ex}{IntegerParam} & \raisebox{1.5ex}{DoubleParam} & Qubit & X & Z & H & T
\\
\hline 
\hline
 
main & 0 & 0 & 2 & 400 & 27800 & 54300 & 55100  \\
\hline
Oracle & 0 & 0 & 0 & 1 & 76 & 137 & 140 \\
\hline
Oracle & 1 & 0 & 0 & 1 & 65 & 130 & 132 \\
\hline
Oracle & 2 & 0 & 0 & 1 & 64 & 142 & 142 \\
\hline
Oracle & 3 & 0 & 0 & 1 & 73 & 134 & 137 \\

\hline 
\end{tabular} 
}
\label{tab:hash_table} 
\end{table} 

\subsection{Timing Analysis}\label{subsec:Timing}
For large quantum problems, simulation on a classical computer is essentially impossible. However, it is useful to have an estimate of the amount of time the algorithm is likely going to take if it were to be scheduled on a quantum computer. Even if the compiler has no knowledge about a hardware implementation's resource constraints, high-level timing analysis can estimate the circuit's critical path length by reordering instructions in order to optimize the logical circuit's length.  For a given sequence of quantum instructions, ScaffCC performs a hierarchical critical path estimation, which involves the scheduling of instructions with the assumption of unbounded quantum resources. The \emph{no-cloning} theorem enforces a data dependency between quantum instructions when they share one or more operands (there is no difference between reads or writes, contrary to classical computing.) Adhering to these dependencies, the critical path timing analysis schedules operations by reordering instructions in \emph{as-soon-as-possible (ASAP)} order. 

Since the quantum program traces can be exceedingly large, we take advantage of modularity to arrive at a critical time estimate. The algorithm proceeds in postorder of the call graph of the program, processing leaf modules before the non-leaf ones. 
The algorithms of Fig.~\ref{fig:crit_path_algo} describe the analysis. It uses a \emph{last\_timestep} table to keep track of the latest timestep in which a qubit was scheduled in an operation. Traversing instructions of a leaf module in program order, a \emph{last\_timestep} table lookup is performed for all operands of an instruction, since each operand may represent a data dependency. This instruction is then scheduled in the earliest timestep possible, resulting in an update in the \emph{last\_timestep} data for its operands. Once all instructions in a module are processed, its \emph{last\_timestep} data is stored, referencing each operand by its argument number for modular analysis. 

For a non-leaf module, the algorithm proceeds in a similar manner, except that when it encounters a module invocation instruction parameterized with qubit arguments, the arguments are treated as its operands. The \emph{last\_timestep} table is examined to determine the earliest timestep the module can be scheduled to start. The values from the \emph{last\_timestep} table of invoked module are added to that of the invoking module, increasing the critical path length for the instruction's operands by the pre-computed value. Once all modules are processed, the highest timestep among all operands is recorded as the critical path of the module

\begin{figure*}[t!]
    \centering
    \begin{subfigure}[t]{\textwidth}
        \centering
        \includegraphics[height=2.7in]{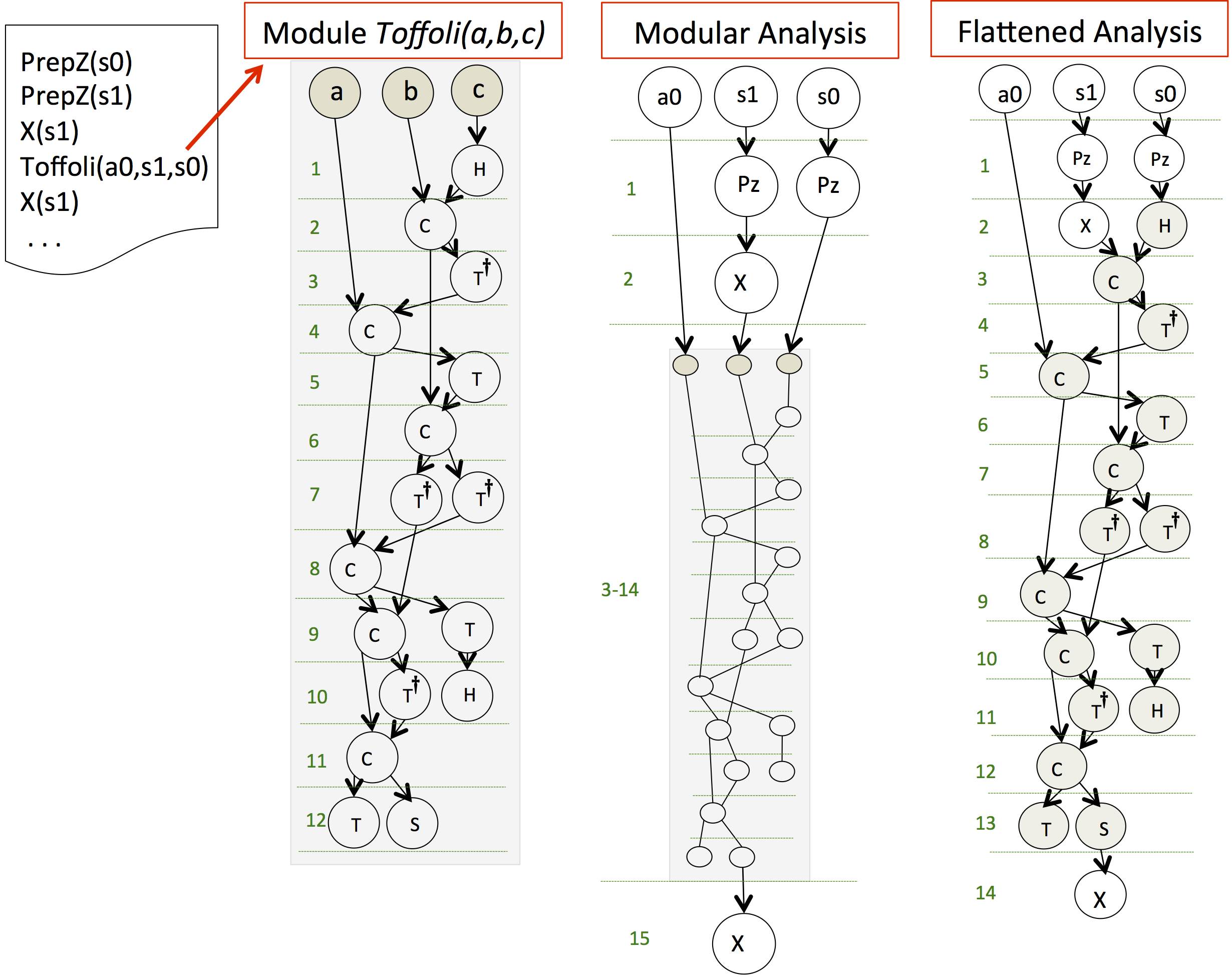}
        \caption{The effect of a modular algorithm for critical path analysis in a toy program (left). Modular analysis improves the runtime of the overall analysis by boxing certain parts of the code. However, the lost parallelism in module boundaries causes longer reported critical paths (center). Remodularization removes box boundaries by inlining some modules into their parent modules, and exposing more parallelism. This tends to increase the code size and is therefore slower, albeit more accurate (right).}
        \label{fig:toffoli}
    \end{subfigure}%
    \\
    \begin{subfigure}[t]{\textwidth}
        \centering
        \includegraphics[height=1.9in]{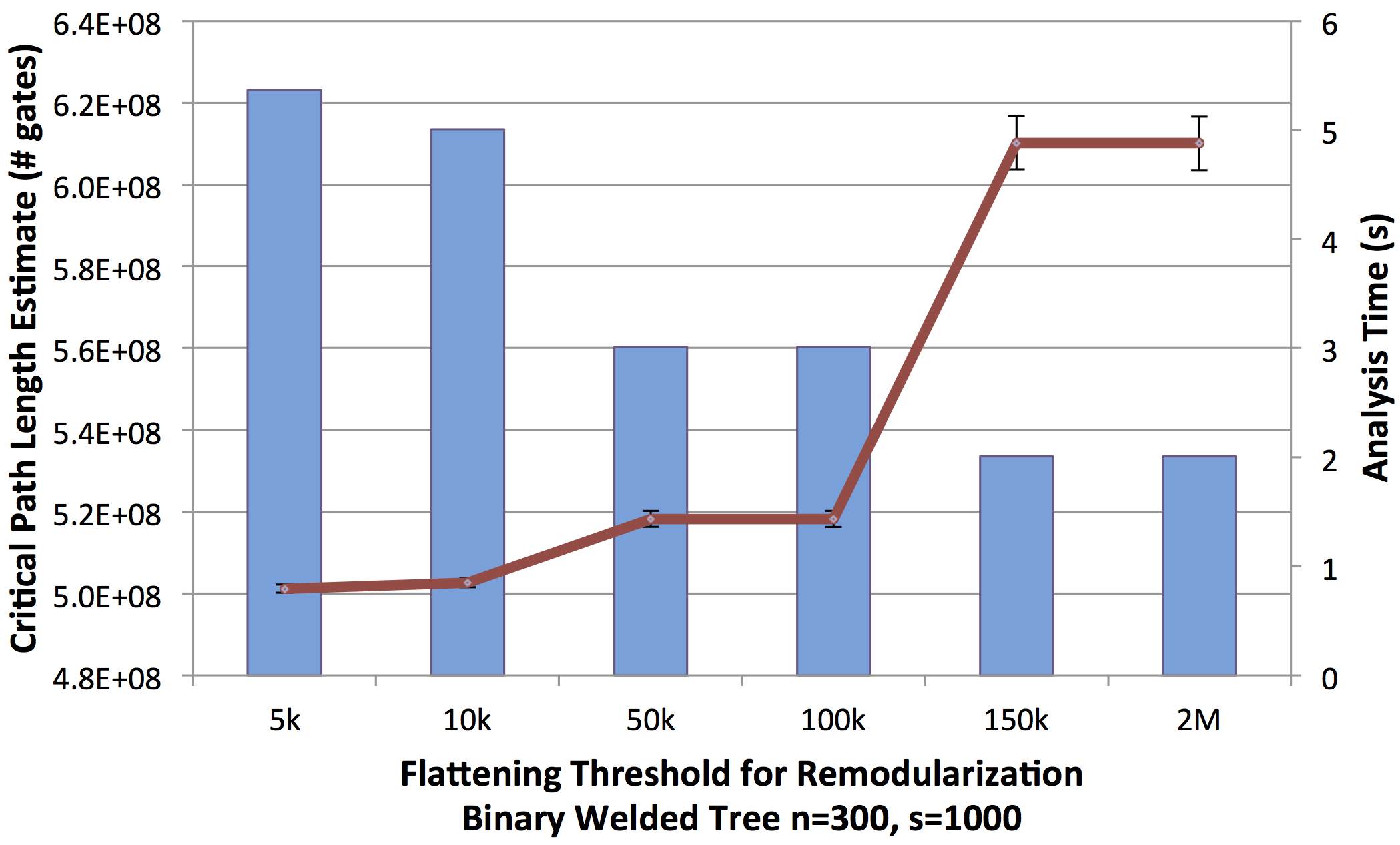}
        \caption{The effect of modularity on circuits' critical time estimate for the case of the BWT(n=300, s=1000) benchmark. More circuit flattening (modular inlining) causes shorter (closer to real) reported critical paths, at the cost of longer analysis time.}
   \label{fig:critical_time}        
    \end{subfigure}
    \caption{Accuracy-Speed Tradeoff in Critical Path Estimation}
\end{figure*}

{\bf Remodularization:} In the analysis of large benchmarks, we use the modularity of a program to avoid repetitive analysis and thereby improve analysis time.
However, this comes at the cost of decreased schedule quality. For example, parallelism between module boundaries can be overlooked in non-flattened sequences of instructions. 
Fig.~\ref{fig:toffoli} depicts this loss of parallelism with an example.
To strike a balance between modularity and optimizability, we perform \emph{remodularization} of the input quantum program. The process involves inlining modules that are too small for optimization, into their respective call sites, and obtaining larger flattened modules. We define a threshold for module size in terms of the number of quantum gates it contains. Informed about module sizes from resource estimation analysis, a remodularization pass in ScaffCC flattens the modules that are smaller than the threshold. Fig.~\ref{fig:critical_time} shows how the estimated critical path gets better with more flattening as more inter-modular parallelism gets discovered.

{\bf Slack-Aware Critical Path Scheduling:} To improve the accuracy of the modular technique, we add the complexity of a small degree of boundary analysis before using pre-determined information from a module. An ASAP schedule, as we have been discussing, is naturally aligned at its top boundary. However, the critical path lengths of all operands in a flat module's schedule may not be equal, creating slacks for some operands at the bottom boundaries. Subsequent operations on these operands may be scheduled in the timesteps that constitute this slack, thereby enabling a shorter schedule. We call such a schedule a \emph{Bottom-Slack Aware} schedule.  To allow the scheduler to account for such slacks, information about the critical path length of each operand in a module must be available. That is, the \emph{last\_timestep} information collected during the scheduling of a module must be provided to the scheduler when scheduling a module composed from it. While this increases the memory requirement of the algorithm slightly, it greatly improves the accuracy of the critical path estimation.

Further improvement is possible if we analyze both the top and the bottom boundaries of a schedule. In this technique, we first create slacks at the top and bottom boundaries by generating a \emph{Center-Aligned} schedule that is densely scheduled at the center than at the boundaries. This allows the scheduler to exploit the boundary slacks to fit modular schedules more tightly into each other. In order to achieve this, our technique relies on a schedule adjustment step on each flat module once its critical path is determined. This step reschedules instructions in the top half of the schedule \emph{as-late-as-possible(ALAP)}, while retaining the instructions in the second half in their ASAP schedule.

\begin{figure}
   \centering
   \includegraphics[width=0.8\linewidth]{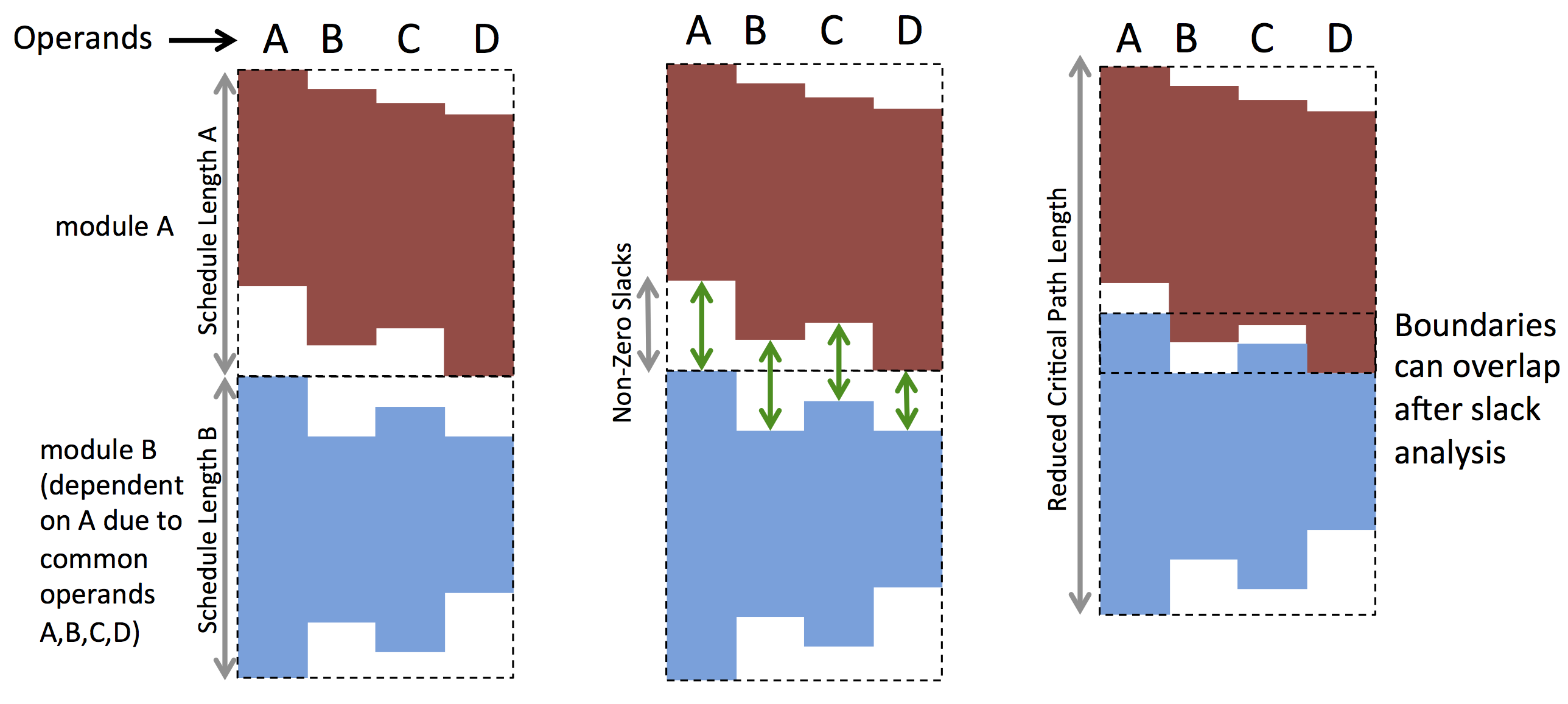}
   \caption{Slack-Aware Timing Analysis: Top and bottom boundary slacks are an opportunity for more fine-grained instruction reordering.}
   \label{fig:slacks}
\end{figure}

\newcommand{\finegrain}{%
\SetAlFnt{\tiny}
\begin{algorithm}[H]
 \For {each flat module $M$}{
 //Step1: ASAP schedule to determine \emph{CriticalPathLength}\\
 //initialize table to hold information about where previous dependencies were 
 //scheduled\\
 \emph{last\_timestep}[$M$][$o_i$] = 0\\
   \For {each instruction $I$ in program order}{
   // Determine earliest timestep in which\\ 
   // the operation can be performed\\
     \For{each operand $o_i$}{
	get \emph{last\_timestep}[$M$][$o_i$]
	}	
	$lt$ = max(\emph{last\_timestep}[$M$][$o_i$])\\
 	Schedule $I$ in timestep $lt$+1\\
	//Update \emph{last\_timestep} table\\
	\For{each operand $o_i$}{
	\emph{last\_timestep}[$M$][$o_i$] = $lt$+1
	}
   }
   //Compute \emph{CriticalPathLength} as length of ASAP schedule\\   
   //Step2: ALAP schedule for top half of critical path schedule\\
   \For {$ts$ = floor(\emph{CriticalPathLength}/2)+1 to \emph{CriticalPathLength} \\
   OR until each qubit $q_i$ in module has been encountered}{
		\For {each instruction $I$ scheduled in $ts$}{
		//find earliest cycle where operand $o_i$ is used\\
		\emph{next\_timestep}[$M$][$q_i$] = earliest cycle $ts$ where $o_i$ is used
		}
}
   
   \For {$ts$ = floor(\emph{CriticalPathLength}/2) to 1}{
		\For {each instruction $I$ scheduled in $ts$}{
		//find latest cycle to schedule each operand $o_i$\\
		$nt$ = min(\emph{next\_timestep}[$o_i$])
		}	
Schedule instruction in timestep $nt$-1\\
//Update \emph{next\_timestep} for each operand\\
\For{each operand $o_i$}{
\emph{next\_timestep}[$M$][$o_i$] = $nt$-1
}
}
\For {each qubit $q_i$ in $M$}{
save \emph{start\_timestep}[$M$][$q_i$]\\
}
}
\end{algorithm}}

\newcommand{\coursegrain}{%
\SetAlFnt{\tiny}
\begin{algorithm}[H]
\For {each non-flat module $M$ in post-order of program's callgraph}{
 //initialize table to hold information about where previous dependencies //were 
 scheduled\\
 \emph{last\_timestep}[$M$][$o_i$] = 0\\
 
   \For {each instruction $I$ in program order}{
   // Determine earliest timestep in which\\ 
   // the operation can be performed\\
     \For{each operand $o_i$}{
	get \emph{last\_timestep}[$M$][$o_i$]
	}	
	$lt$ = max(\emph{last\_timestep}[$M$][$o_i$])
   }
   \eIf {$I$ is a quantum gate}{
	Schedule $I$ in timestep $lt$+1\\
	//Update \emph{last\_timestep} table\\
	\For{each operand $o_i$}{
	\emph{last\_timestep}[$M$][$o_i$] = $lt$+1
}
	}
	{
	//$I$ is a module invocation\\
	//Compute slack for each operand
	\For{each operand $o_i$}{
	\emph{slack}[$o_i$] = $lt$+\emph{start\_timestep}[$I$][$o_i$] - \emph{last\_timestep}[$M$][$o_i$]
	}
	$least\_slack$ = min(\emph{slack}[$o_i$])\\
	Schedule $I$ to start executing from timestep ($lt$ + 1 - \emph{least\_slack})\\
	//Update \emph{last\_timestep} table\\
\For{each operand $o_i$}{
\emph{last\_timestep}[$M$][$o_i$] = $lt$ + 1 - \emph{leastslack} + \emph{last\_timestep}[$I$][$o_i$]
}
}
 }
\end{algorithm}}

\begin{figure}
\centering
    \begin{subfigure}{.5\textwidth}
        \finegrain
	\caption{Center-Aligned scheduling of leaf modules.}
	 \label{algo:ca_schedule}    	   
    \end{subfigure}
    \begin{subfigure}{.5\textwidth}
        \coursegrain
	\caption{Composing overall schedule in ASAP manner.}
	 \label{algo:asap_reschedule}         
	 \end{subfigure}\\
\caption{Fine-Grained (a) and Coarse-Grained (b) algorithms for scheduling leaf and non-leaf modules respectively. The fine-grained schedule uses a center-aligned approach to create the most slack at the top and bottom boundaries, and the coarse-grained algorithm utilizes the remaining slots in the slacks to compress the overall schedule as much as possible.}
\label{fig:crit_path_algo}
\end{figure}

\begin{figure*}[t!]
    \centering
    \begin{subfigure}[t]{0.5\textwidth}
        \centering
        \includegraphics[height=1.4in]{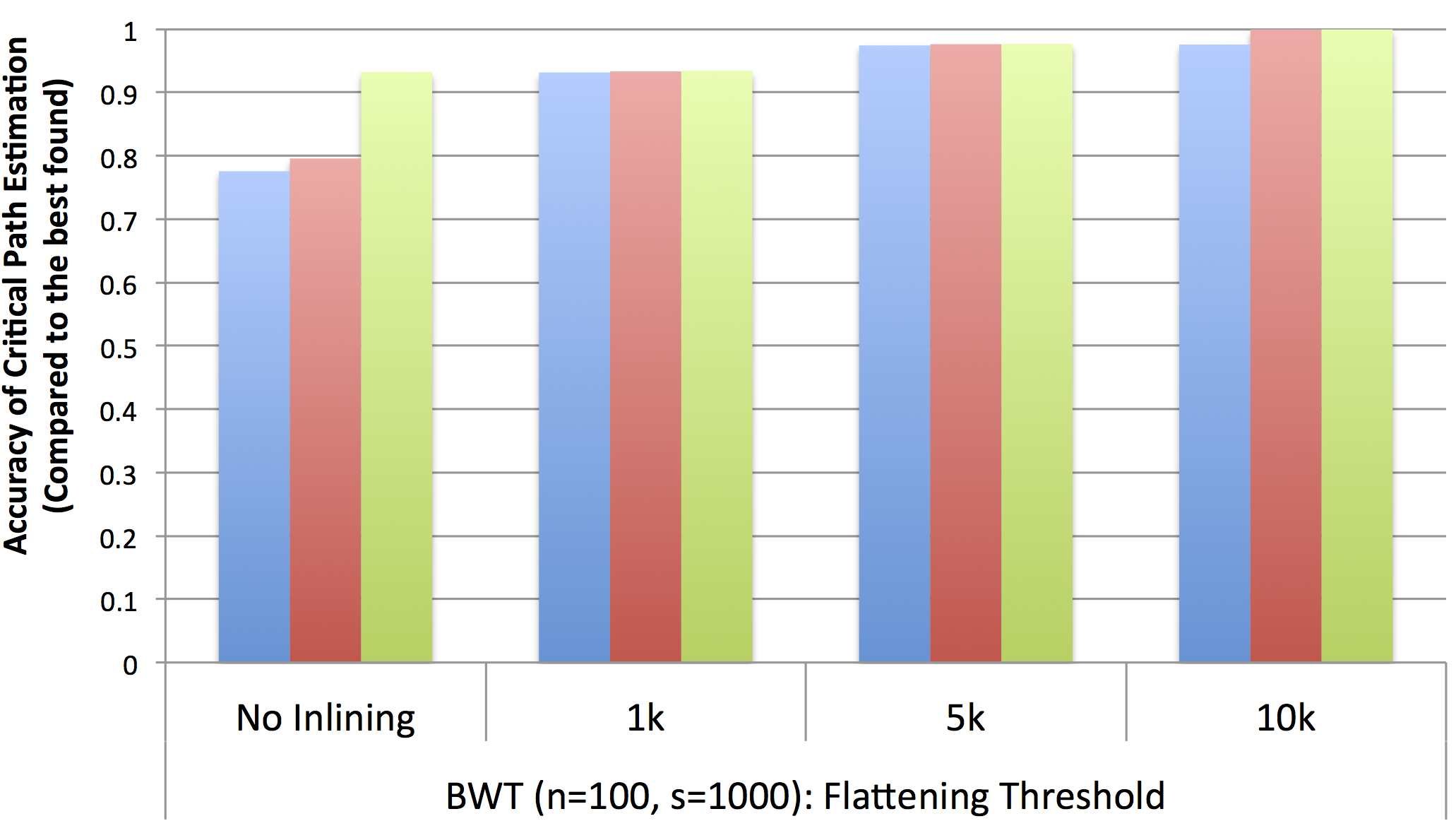}
        \caption{}
        \label{fig:bwt_crit_path}
    \end{subfigure}%
    ~ 
    \begin{subfigure}[t]{0.5\textwidth}
        \centering
        \includegraphics[height=1.4in]{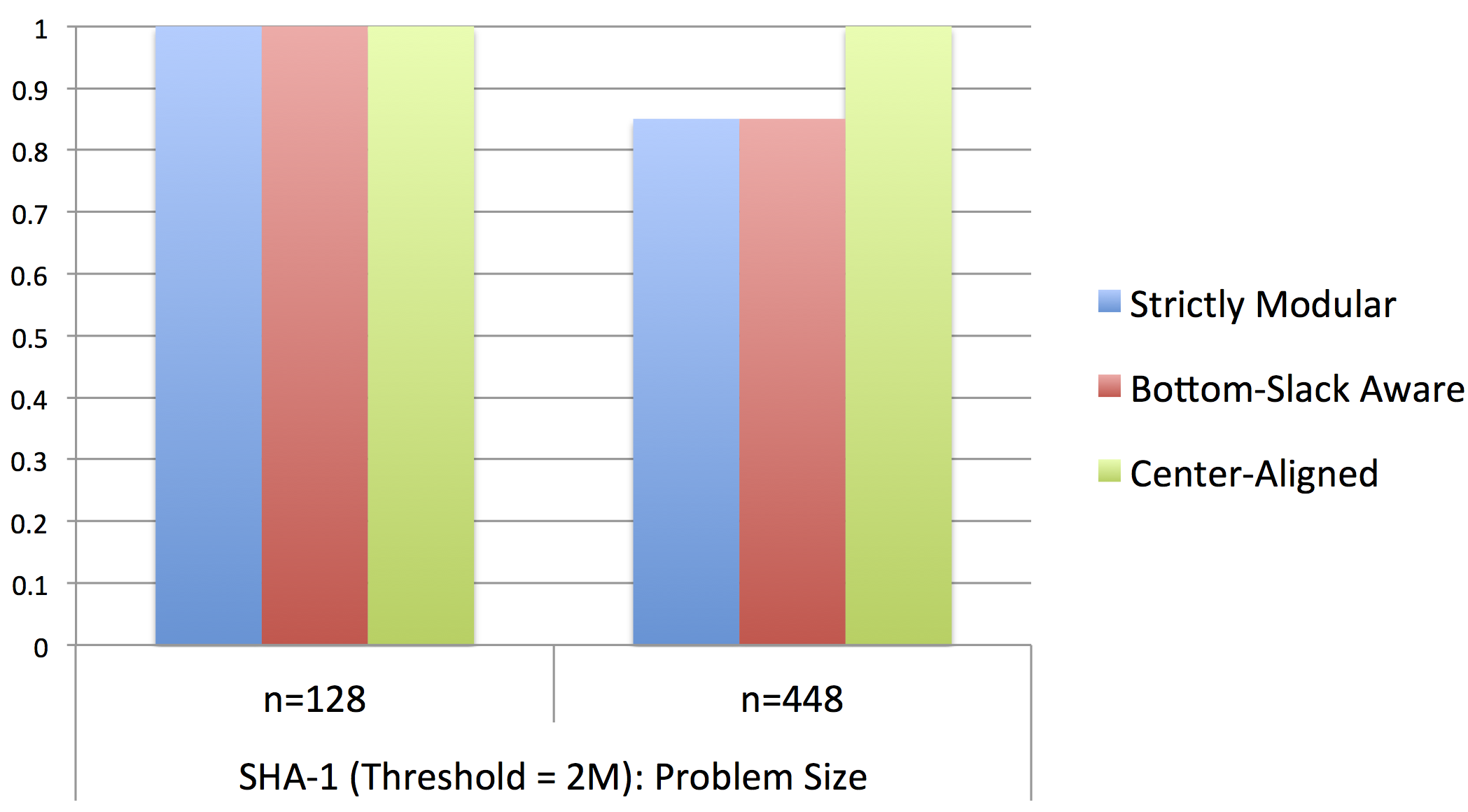}
        \caption{}
           \label{fig:sha1_crit_path}        
    \end{subfigure}
    \caption{The effect of increasing thresholds and increasing timing analysis complexity: More flattening and higher-complexity analysis result in higher accuracies (a). In some instances the effect of higher-complexity analysis can be masked with mere flattening (b - left), but flattening can only be done up to a certain threshold without causing huge increase in code size. In these cases, which occur in larger problems, we can rely on the center-aligned scheduling method for recovering some of the lost parallelism (b-right).}
    \label{fig:critical_path_comparison}
\end{figure*}

Figure~\ref{fig:critical_path_comparison} shows the effect of these three different scheduling algorithms for estimating the critical path. It can be seen that the new slack-aware estimation methods, as well as flattening, can work hand-in-hand to discover the parallelism of the code which may be lost in its modular design.

\section{Related Work}
\label{sec:RelatedWork}
This paper is an extension of previous work that introduced ScaffCC \cite{CF}, with a broader set of benchmarks and techniques as well as more in-depth analysis.

Many previous works on high-level quantum programming have focused on the design of programming languages rather than compiler design. Programming languages based on C~\cite{Omer} and Haskell~\cite{qio_monad,Quipper} have been proposed for quantum computing applications
specifically to facilitate development of correct quantum algorithms. In contrast, ScaffCC is a compiler effort that studies and develops compiler strategies for efficient quantum compilation and analysis. In particular ScaffCC differentiates itself through two objectives: generation of tractable quantum assembly code that is also amenable to aggressive low-level optimizations; and logical program analyses at full program scale. Similar to Quipper as proposed by Green et. al.~\cite{Quipper}, the ScaffCC compiler handles program scale by making heavy use of modularization. Additionally, ScaffCC recognizes the implications of the degree of modularity on both efficiency and quality of compilation, and presents scalable techniques to achieve both.

Moreover, this paper presents a first study in the trade-off between modular inlining and critical-time estimation accuracy.
Although other papers envision the prospect of the QASM language being an extension of conventional classical assembly languages extended with a quantum instruction set \cite{Elhoushi, Oskin-computer}, to the best of our knowledge none have implemented large circuits using this format and studied the trade-offs between manageability and optimizability.

Our work in CTQG is a pioneering effort to create a C-language-like reversible logic compiler for quantum circuits. Although other tools exist which work on small circuits and try to find \emph{optimal} decompositions into reversible gates~\cite{revlib}, similar to~\cite{Quipper} our compiler scales to arbitrary size problems but also includes a state-of-the-art algorithm for synthesis of integer arithmetic which generates no ancilla signals, a well-optimized library of fixed-point analytic functions and an automatic ancilla manager which significantly reduces the use of ancillas in comparison to a brute-force approach.

Previous work has enabled resource analysis as part of algorithm development~\cite{Quipper, Omer}. ScaffCC expands its analysis toolbox with other useful analyses such as entanglement and timing analysis. The analysis framework can be easily extended further. For example, Metodi et al.~\cite{MetodiCompiler} propose a useful reliability analysis in circuits, whose results can be compared with the reliability goal of the hardware and used to determine circuit locations in need of error correction. Techniques for exact entanglement analysis have been previously proposed in~\cite{Perdrix,Prost}. Perdrix~\cite{Perdrix} has developed typed language extensions for abstract interpretation of entanglements in quantum data arrays, while Prost and Zerrari~\cite{Prost} have proposed formal semantics for identifying entanglements in higher order functions. ScaffCC performs a conservative and modular entanglement analysis at a purely logical level. This is intended to aid both design and debugging of quantum algorithms, which benefit from an understanding of where entanglements are potentially created and removed in a program.
Furthermore, Schuchman and Vijaykumar~\cite{Compute_Uncompute} identify a program transformation which exploits parallelism between computation and uncomputation portions of a program, albeit at the cost of increased qubits. This transformation can easily be added to ScaffCC due to its tracking of uncompute regions.

\section{Conclusion}
This paper has examined the issues concerning the high-level compilation of quantum circuits. We showed the possibility of compiling large-scale applications, with the applicability of some previous classical techniques and also opportunities for exploiting the dual classical-quantum nature of programs for keeping the compilation process tractable. Methods for program correctness checking as well as a novel approach to reversible-logic synthesis were also proposed. We also presented a detailed discussion of methods for estimating circuit critical paths, and explained how they can be optimized in the face of increasing code size. The trade-off between optimality and speed in this analysis was evaluated. These discussions form a stepping stone towards efficient mapping of quantum algorithms onto physical quantum computers in the future.

The ScaffCC software is available for download at: \url{https://github.com/ajavadia/ScaffCC}.

\section{Acknowledgements}
The authors acknowledge funding from the NSF PHY-1415537 grant.

The quantum benchmarks used in this paper were developed by the following researchers: Amlan Chakrabarti (Princeton), John Black (Colorado), Oana Catu and Daniel Kudrow (UCSB), Mohammad Javad Dousti (USC), Chen-Fu Chiang (Sherbrooke).

\bibliography{references}

\end{document}